

\font\twelverm=cmr10 scaled 1200    \font\twelvei=cmmi10 scaled 1200
\font\twelvesy=cmsy10 scaled 1200   \font\twelveex=cmex10 scaled 1200
\font\twelvebf=cmbx10 scaled 1200   \font\twelvesl=cmsl10 scaled 1200
\font\twelvett=cmtt10 scaled 1200   \font\twelveit=cmti10 scaled 1200
\font\twelvesc=cmcsc10 scaled 1200
\skewchar\twelvei='177   \skewchar\twelvesy='60
\def\twelvepoint{\normalbaselineskip=12.4pt
  \medskipamount=7.2pt plus2.4pt minus2.4pt
  \def\rm{\fam0\twelverm}          \def\it{\fam\itfam\twelveit}%
  \def\sl{\fam\slfam\twelvesl}     \def\bf{\fam\bffam\twelvebf}%
  \def\mit{\fam 1}                 \def\cal{\fam 2}%
  \def\tt{\twelvett}
  \def\sc{\twelvesc}
  \def\nullspace{\nulldelimiterspace=0pt \mathsurround=0pt }
  \def\big##1{{\hbox{$\left##1\vbox to 10.2pt{}\right.\nullspace$}}}
  \def\Big##1{{\hbox{$\left##1\vbox to 13.8pt{}\right.\nullspace$}}}
  \def\bigg##1{{\hbox{$\left##1\vbox to 17.4pt{}\right.\nullspace$}}}
  \def\Bigg##1{{\hbox{$\left##1\vbox to 21.0pt{}\right.\nullspace$}}}
  \textfont0=\twelverm   \scriptfont0=\tenrm   \scriptscriptfont0=\sevenrm
  \textfont1=\twelvei    \scriptfont1=\teni    \scriptscriptfont1=\seveni
  \textfont2=\twelvesy   \scriptfont2=\tensy   \scriptscriptfont2=\sevensy
  \textfont3=\twelveex   \scriptfont3=\twelveex  \scriptscriptfont3=\twelveex
  \textfont\itfam=\twelveit
  \textfont\slfam=\twelvesl
  \textfont\bffam=\twelvebf \scriptfont\bffam=\tenbf
  \scriptscriptfont\bffam=\sevenbf
  \normalbaselines\rm}

\def\beginlinemode{\endmode
  \begingroup\parskip=0pt \obeylines\def\\{\par}\def\endmode{\par\endgroup}}
\def\beginparmode{\endmode \begingroup \def\endmode{\par\endgroup}}
\let\endmode=\par
{\obeylines\gdef\
{}}
\def\singlespace{\baselineskip=\normalbaselineskip}
\def\oneandahalfspace{\baselineskip=\normalbaselineskip \multiply\baselineskip
     by 3 \divide\baselineskip by 2}
\def\doublespace{\baselineskip=\normalbaselineskip \multiply\baselineskip by 2}
\newcount\firstpageno \firstpageno=2
\footline={\ifnum\pageno<\firstpageno{\hfil}\else{\hfil\twelverm\folio\hfil}\fi}
\let\rawfootnote=\footnote		
\def\footnote#1#2{{\rm\parindent=20pt\singlespace\hang
  \rawfootnote{#1}{\tenrm#2\hfill\vrule height 0pt depth 6pt width 0pt}}}
\def\raggedcenter{\leftskip=4em plus 12em \rightskip=\leftskip
  \parindent=0pt \parfillskip=0pt \spaceskip=.3333em \xspaceskip=.5em
  \pretolerance=9999 \tolerance=9999 \hyphenpenalty=9999 \exhyphenpenalty=9999
}
\parskip=\medskipamount \twelvepoint \overfullrule=0pt
\def\title {\null\vskip 3pt plus 0.3fill \beginlinemode
   \doublespace \raggedcenter \bf}
\def\author{\vskip 3pt plus 0.3fill \beginparmode \raggedcenter \sc}
\def\affil{\vskip 3pt plus 0.1fill \beginlinemode
   \oneandahalfspace \raggedcenter \sl}
\def\abstract {\vskip 3pt plus 0.3fill \beginparmode
   \oneandahalfspace \narrower ABSTRACT:~~}
\def\endtitlepage{\vfill\eject\beginparmode}
\def\subhead#1{\vskip 0.25truein{\raggedcenter #1 \par} \nobreak
   \vskip 0.25truein\nobreak}
\def\references{\subhead{References}
   \frenchspacing \parindent=0pt \leftskip=0.8truecm \rightskip=0truecm
   \parskip=4pt plus 2pt \everypar{\hangindent=\parindent}}
\def\endreferences{\beginparmode}
\def\refstylenp{
  \gdef\refto##1{~[##1]}                                
  \gdef\refis##1{\indent\hbox to 0pt{\hss[##1]~}}     	
  \gdef\citerange##1##2##3{~[\cite{##1}--\setbox0=\hbox{\cite{##2}}\cite{##3}]}
  \gdef\journal##1, ##2, ##3, ##4{{\sl##1} {\bf ##2} (##3) ##4} }
\def\cmp{\journal Comm. Math. Phys.}
\def\np{\journal Nucl. Phys.}
\def\endit{\endmode\vfill\supereject\end}
\def\half{{\textstyle{ 1\over 2}}}
\def\ts{\textstyle}
\def\gtwid{\mathrel{\raise.3ex\hbox{$>$\kern-.75em\lower1ex\hbox{$\sim$}}}}
\def\ucsb{Department of Physics\\University of California\\
          Santa Barbara, CA 93106}
\def\a{\alpha}
\def\b{\beta}
\def\d{\delta}
\def\r{\rho}
\def\s{\sigma}
\def\gev{{\rm \,Ge\kern-0.125em V}}
\def\ket#1{|{#1}\rangle}
\def\bra#1{\langle{#1}|}
\def\braoket#1#2#3{\langle{#1}|{#2}|{#3}\rangle}
\def\Tr{\mathop{\rm Tr}\nolimits}
\def\hsl{H\kern-.7em\raise.15ex\hbox{$/$}\>}

\def\sbar{\overline\sigma}
\def\bd{b^\dagger}
\def\Qa{Q^{\vphantom{\dagger}}_{\a\vphantom{\dot\a}}}
\def\Qad{Q^\dagger_{\dot\a}}
\def\bi{b^{\vphantom{\dagger}}_i}
\def\bid{b^\dagger_i}
\refstylenp
\catcode`@=11
\newcount\r@fcount \r@fcount=0
\newcount\r@fcurr
\immediate\newwrite\reffile
\newif\ifr@ffile\r@ffilefalse
\def\w@rnwrite#1{\ifr@ffile\immediate\write\reffile{#1}\fi\message{#1}}
\def\writer@f#1>>{}
\def\referencefile{
  \r@ffiletrue\immediate\openout\reffile=\jobname.ref%
  \def\writer@f##1>>{\ifr@ffile\immediate\write\reffile%
    {\noexpand\refis{##1} = \csname r@fnum##1\endcsname = %
     \expandafter\expandafter\expandafter\strip@t\expandafter%
     \meaning\csname r@ftext\csname r@fnum##1\endcsname\endcsname}\fi}%
  \def\strip@t##1>>{}}

\def\citeall#1{\xdef#1##1{#1{\noexpand\cite{##1}}}}
\def\cite#1{\each@rg\citer@nge{#1}}	
\def\each@rg#1#2{{\let\thecsname=#1\expandafter\first@rg#2,\end,}}
\def\first@rg#1,{\thecsname{#1}\apply@rg}	
\def\apply@rg#1,{\ifx\end#1\let\next=\relax
\else,\thecsname{#1}\let\next=\apply@rg\fi\next}
\def\citer@nge#1{\citedor@nge#1-\end-}	
\def\citer@ngeat#1\end-{#1}
\def\citedor@nge#1-#2-{\ifx\end#2\r@featspace#1 
  \else\citel@@p{#1}{#2}\citer@ngeat\fi}	
\def\citel@@p#1#2{\ifnum#1>#2{\errmessage{Reference range #1-#2\space is bad.}%
    \errhelp{If you cite a series of references by the notation M-N, then M and
    N must be integers, and N must be greater than or equal to M.}}\else%
 {\count0=#1\count1=#2\advance\count1
by1\relax\expandafter\r@fcite\the\count0,%
  \loop\advance\count0 by1\relax
    \ifnum\count0<\count1,\expandafter\r@fcite\the\count0,%
  \repeat}\fi}
\def\r@featspace#1#2 {\r@fcite#1#2,}	
\def\r@fcite#1,{\ifuncit@d{#1}
    \newr@f{#1}%
    \expandafter\gdef\csname r@ftext\number\r@fcount\endcsname%
                     {\message{Reference #1 to be supplied.}%
                      \writer@f#1>>#1 to be supplied.\par}%
 \fi%
 \csname r@fnum#1\endcsname}
\def\ifuncit@d#1{\expandafter\ifx\csname r@fnum#1\endcsname\relax}%
\def\newr@f#1{\global\advance\r@fcount by1%
    \expandafter\xdef\csname r@fnum#1\endcsname{\number\r@fcount}}
\let\r@fis=\refis			
\def\refis#1#2#3\par{\ifuncit@d{#1}
   \newr@f{#1}%
   \w@rnwrite{Reference #1=\number\r@fcount\space is not cited up to now.}\fi%
  \expandafter\gdef\csname r@ftext\csname r@fnum#1\endcsname\endcsname%
  {\writer@f#1>>#2#3\par}}
\let\r@ferences=\references
\def\references{\r@ferences\def\endmode{\r@ferr\par\endgroup}}
\let\endr@ferences=\endreferences
\def\endreferences{\r@fcurr=0
  {\loop\ifnum\r@fcurr<\r@fcount
    \advance\r@fcurr by 1\relax\expandafter\r@fis\expandafter{\number\r@fcurr}%
    \csname r@ftext\number\r@fcurr\endcsname%
  \repeat}\gdef\r@ferr{}\endr@ferences}
\def\reftorange#1#2#3{\citerange{#1}{#2}{#3}}

\let\r@fend=\endpaper\gdef\endpaper{\ifr@ffile
\immediate\write16{Cross References written on []\jobname.REF.}\fi\r@fend}
\catcode`@=12
\citeall\refto

\singlespace
\rightline{hep-th/9206056}
\rightline{UCSBTH--92--22}
\rightline{Revised July 1993}
\doublespace
\title  IS PURITY ETERNAL$\,$?
\author Mark Srednicki\footnote{*}{Electronic address:
mark@tpau.physics.ucsb.edu}
\affil \ucsb
\abstract  Phenomenological and formal restrictions on the evolution of
pure into mixed states are discussed.  In particular, it is argued that,
if energy is conserved,
loss of purity is incompatible with the weakest possible form of
Lorentz covariance.
\endtitlepage
\baselineskip=16pt

\subhead{1. Introduction}

In 1983, Hawking proposed that, due to quantum gravitational effects, a
scattering process might result in an initial pure state becoming a final
mixed state\refto{hawking83}.
This idea (which had been suggested earlier as a phenomenological possibility
worthy of investigation\refto{marinov74})
is clearly one of great intellectual power.
It would represent a truly fundamental modification of the laws of physics.

Nevertheless, Hawking's proposal quickly received criticism.
It was noted\refto{gross84,ehns84} that conservation laws become decoupled
from symmetry principles,
so that (for example) rotation invariance would no longer imply conservation
of angular momentum.  Furthermore, actual calculations in models
which admit systematic approximations showed no hint of the decay of
purity\refto{gross84} (see, however, ref.\refto{hawking84}).
Also, since quantum mechanics is well tested in a variety
of systems, corrections to the generalized hamiltonian would have to
be quite small, with eigenvalues of order $10^{-21}\gev$ or less\refto{ehns84}.

By far the most severe attack was leveled by Banks, Peskin, and Susskind
(BPS)\refto{bps84},
who concluded that conservation of energy and momentum could not
be preserved without losing locality.  Since then, however, the
prevailing notion of the meaning of locality has weakened somewhat,
due to the discovery of the wormhole phenomenon\refto{gs88a}.
Wormholes violate locality, but the phenomenological results turn out to be
remarkably benign\reftorange{gs88b}{coleman88a}{ksb89}.
It therefore seems appropriate to reexamine Hawking's proposal,
to see whether or not the breakdown of locality noted by BPS is truly
objectionable, and to see what other constraints, if any, may be placed on
the decay of purity.

Let us briefly review some of the basic formalism for a theory
in which pure states can evolve to mixed states; for more details,
see refs.\refto{marinov74,ehns84,bps84}.  Such a theory must be formulated
in terms of a density matrix $\r$ rather than a state $\ket\psi$.
The eigenvalues of $\r$ represent probabilities, and so
must be real, positive or zero, and sum to one (assuming normalizable states).
Thus $\rho$ must be hermitian and satisfy $\Tr\rho=1$.
The state is said to be pure if $\r=\ket\psi\bra\psi$ for some Hilbert space
vector $\ket\psi$; in this case $\Tr\r^2=\Tr\r=1$.  If $\Tr\rho^2<1$,
then $\r$ cannot be written as $\ket\psi\bra\psi$ for any $\ket\psi$, and
the state is said to be mixed.
The usual evolution equation for $\r$ is $\dot\rho=-i[H,\r]$, and this
leaves both $\Tr\r$ and $\Tr\r^2$ constant in time.
To allow loss of purity, we must modify this equation.
Following refs.\refto{marinov74,ehns84},
we assume that the new evolution equation for $\rho$ is still linear,
and still first order in time derivatives:
$$\dot\rho_{ab}=\hsl_{ab}{}^{cd}\,\rho_{dc}\;.  \eqno(1)$$
The generalized hamiltonian $\hsl$ must be constrained to
preserve the hermiticity, positivity, and trace of $\rho$.
As shown by BPS, the most general equation preserving $\r^\dagger=\r$ and
$\Tr\r=1$ is
$$\dot\r=-i[H,\r] -\half g_{\a\b}\bigl( Q^\a Q^\b\r+\r Q^\a Q^\b
                                       -2Q^\b\r Q^\a\bigr)\;, \eqno(2)$$
where the $Q$'s are any hermitian operators other than the identity,
and $g_{\a\b}$ is a hermitian matrix of coupling constants.
The hamiltonian $H$ is unambiguously defined
as the operator appearing in the commutator term.
BPS also showed that a sufficient condition for $\r$ to remain positive is
that the eigenvalues of $g_{\a\b}$ be positive or zero.

Acceptable phenomenology leads us to
demand that the familiar conservation laws
be satisfied.  Energy conservation, for example, requires that
$\Tr f(H)\dot\r=0$,
where $f(H)$ is any smooth function of the hamiltonian $H$.
BPS have shown that,
if $g_{\a\b}$ has nonnegative eigenvalues and is, in addition,
real and symmetric, then
conservation of energy requires that $H$ commute with each of the $Q$'s.
(If $g_{\a\b}$ does not satisfy all these conditions,
there may be other possibilities; this will be discussed later.)
In a quantum field theory, there are very
few operators available which commute with the hamiltonian, and all of them
are global: that is, integrals over all space of a local
density.\footnote*{In a gauge theory, we have the local generators
of the gauge symmetry; these annihilate all the physical states, however,
and so would vanish in eq.$\,$(2).}  We always
have at our disposal the hamiltonian $H$ and the total momentum operator
$\vec P$ to press into service as $Q$'s.  We may also have global charges
corresponding to conserved quantities like baryon number.
Is eq.$\,$(2), with the $Q$'s chosen from this list, a viable possibility?

BPS concluded that the answer is {\it no}, due to a breakdown of locality.
In particular, they state that cluster decomposition will no longer hold.
Given the density matrix $\r$ at some
fixed time, and localized operators $A(\vec x\,)$ and $B(\vec y\,)$,
cluster decomposition states that
$$\Tr\bigl[\r A(\vec x\,)B(\vec y\,)\bigr]
                    \simeq\Tr\bigl[\r A(\vec x\,)\bigr]
                          \Tr\bigl[\r B(\vec y\,)\bigr]      \eqno(3)$$
when $|\vec x-\vec y\,|$ becomes large.
The precise meaning of ``large'' depends on the theory, and on $\r$.
In ordinary field theory, eq.$\,$(3) is usually quoted as a property of the
vacuum density matrix $\r_0=\ket{0}\bra{0}$.
If the lightest particle in the theory has
mass~$m$, and if we take $\r=\r_0$, then
``large'' means much bigger than $m^{-1}$.
But if, for example, $\r$ represents
a pure state consisting of one or more particles whose wave functions are
localized near the origin but coherent over a region of linear size
$a\gg m^{-1}$, then ``large'' means much bigger than $a$.
The question is, if eq.$\,$(3) is valid for
some initial density matrix $\r_{\rm init}$, and
for $|\vec x-\vec y\,|$ greater than some length $L$,
will this still be true at later times?

If we assume that energy is conserved, then pure energy eigenstates
like $\r_0$ do not evolve in time, and so there is no modification of the
cluster property for $\r_{\rm init}=\r_0$.
If we consider instead a $\r_{\rm init}$ corresponding to
a localized excited state, we must ask how quickly
its wave packet spreads out as time evolves.
Even in ordinary field theory,
wave packets spread out, and the minimum value of $|\vec x-\vec y\,|$ for which
eq.$\,$(3) holds will grow with time.  It turns out that eq.$\,$(2)
{\it does} cause wave packets to spread out a bit faster than
they otherwise would,
but the effect is not dramatic; certainly it does not qualify as a violation
of cluster decomposition.  However, if we put some particles at {\it both}
$\vec x$ and $\vec y$, then extra correlations {\it do} develop with time.
For free nonrelativistic particles,
these remain small at all times, and actually disappear at late times.
This is discussed in detail in sect.~2.
We conclude that the nonlocality noted by BPS, while present, is in fact
difficult to detect, and that eq.$\,$(2) cannot be dismissed because of it.

One may still question whether or not eq.$\,$(2), with global $Q$'s,
has any reasonable chance to arise as the low energy limit of a more
fundamental theory.  I know of no such theory, but there is a strong passing
resemblance between eq.$\,$(2) and the effective theories that arise due to
wormholes\reftorange{gs88a}{gs88b,coleman88a}{ksb89} (which were originally
thought to cause purity to decay).  In wormhole theory, the usual field
theory action $S_0=\int d^4x\,{\cal L}$ is modified to\refto{ksb89}
$$S=S_0+g_{\a\b}Q^\a Q^\b+\ldots      \eqno(4)$$
where $Q^\a=\int d^4x\,q^\a(x)$, $q^\a(x)$ is some local operator,
and the ellipses stand for higher powers
of $Q$'s.  It seems not totally unreasonable to suppose that there could
exist a variant of wormhole theory which ultimately results in eq.$\,$(2)
rather than eq.$\,$(4).  Of course, there are some important differences
between eq.$\,$(2) and eq.$\,$(4): in eq.$\,$(4), the $Q$'s are {\it four}
dimensional integrals, and they modify the action itself.  They therefore do
not lead to violations of quantum mechanics, but do look like they would
produce severe violations of locality.  As is now well known, this is not the
case.  Similarly, the apparent violations of locality in eq.$\,$(2) are not as
drastic as they appear to be at first glance.

If eq.$\,$(2) does not lead to serious violations of locality, and cannot be
immediately rejected on grounds of implausibility, it becomes necessary
to reexamine its viability.  That is the purpose of this paper.
Phenomenological constraints,
including the anomalous spreading of wave packets,
are examined in sect.~2.  Formal restrictions
are discussed in sect.~3, where it is found that eq.$\,$(2) is incompatible
with the weakest possible form of Lorentz covariance (if energy is conserved).
This is the main conclusion of this paper.
However, a possible loophole in the argument is presented in sect.~4,
and conclusions in sect.~5.

\subhead{2. Phenomenological constraints}

As noted in the introduction, the question of the locality of eq.$\,$(2)
requires us to study the spreading of wave packets with time.  Rather than
immediately attacking this problem within quantum field theory, we will warm up
with some examples from one-dimensional, nonrelativistic quantum mechanics that
contain the essential physics.  Let us first consider a harmonic oscillator of
unit mass and frequency (so that the hamiltonian is $H=a^\dagger a+\half$)
which is in a coherent state.  At time $t=0$, the state is specified by an
arbitrary complex parameter $\beta=|\beta|e^{i\phi}$:
$$\ket{\beta,0}=e^{-|\beta|^2/2}e^{\beta a^\dagger}\ket{0}\;.   \eqno(5)$$
It is well known that at later
times the state is still coherent, but with a different value of $\beta$:
$$\ket{\beta,t}=e^{-it/2}\ket{\beta e^{-it},0}\;.   \eqno(6)$$
Furthermore, the probability to find the oscillator at position $x$ at
time $t$ is given by
$$\braoket{x}{\rho(\beta,t)}{x} = {\textstyle{1\over\sqrt\pi}}
                 e^{-\left[x-\sqrt2\,|\beta|\cos(t-\phi)\right]^2}   \eqno(7)$$
where $\rho(\beta,t)=\ket{\beta,t}\bra{\beta,t}$ is the density matrix at
time $t$.  Thus, the oscillator's wave packet just moves back and forth
without changing shape, its center following a classical trajectory
with maximum amplitude $\sqrt2\, |\beta|$.

Now consider modifying the quantum equation of motion to the general form of
eq.$\,$(2).  In order to conserve energy, the $Q$'s must commute with~$H$. For
simplicity, we will take a single $Q$ and set it equal to $H$, and let the
corresponding value of $g$ be real and positive.  To compute the density matrix
at time $t$, we first expand the initial state $\ket{\beta,0}$ into energy
eigenstates:
$$\ket{\beta,0}=e^{-|\beta|^2/2}\sum_{n=0}^\infty{\beta^n\over(n!)^{1/2}}
                 \ket{n}\;,   \eqno(8)$$
where $\ket{n}=(n!)^{-1/2}(a^\dagger)^n\ket{0}$
is a normalized energy eigenstate with energy $n+\half$.
The time evolution of a density matrix element $\ket{n}\bra{m}$
in this basis is simple, and results in
$$\rho(\beta,t)=e^{-|\beta|^2}\sum_{n,m=0}^\infty{\beta^n\beta^{*m}
    \over(n!m!)^{1/2}}\,e^{-i(n-m)t-g(n-m)^2t/2}\,\ket{n}\bra{m}\;.\eqno(9)$$
The probability of finding the oscillator at position $x$ at time $t$ is again
$\langle x|\rho(\beta,t)|x\rangle$; let us study the behavior of
this function as $t\to\infty$.
Looking at $\rho(\beta,t)$, we see that its off-diagonal components decay away
at late times, and it approaches
$$\rho(\beta,\infty)=e^{-|\beta|^2}\sum_{n=0}^\infty{|\beta|^{2n}
                    \over n!}\,\ket{n}\bra{n}\;.           \eqno(10)$$
Furthermore, this final state is independent of the specific choice of the
$Q$'s as long as they all commute with $H$.
To compute $\langle x|\rho(\beta,\infty)|x\rangle$ explicitly, note that
the off-diagonal components of $\rho(\beta,t)$ can also be removed by
averaging over the phase of $\beta$:
$$\rho(\beta,\infty)={1\over2\pi}\int_0^{2\pi}d\phi\,\rho(\beta,t)\;.\eqno(11)$$
Since the result is independent of time, we may as well use
$\rho(\beta,0)=\ket{\beta,0}\bra{\beta,0}$
on the right hand side.  Then, taking the expectation value in a position
eigenstate and using eq.$\,$(7) at $t=0$, we find
$$\langle x|\rho(\beta,\infty)|x\rangle={1\over2\pi^{3/2}}\int_0^{2\pi}d\phi\,
                 e^{-[x-\sqrt2\,|\beta|\cos\phi]^2}\;.  \eqno(12)$$
We thus see that the oscillator forgets its starting point, but
conservation of energy forces it to remember its
maximum amplitude $\sqrt2\, |\beta|$.  If we make a change of integration
variable in eq.$\,$(12), we get
$$\langle x|\rho(\beta,\infty)|x\rangle
    = \int_{-\sqrt2\,|\beta|}^{+\sqrt2\,|\beta|}dy\,
      \left[ {\textstyle{1\over\sqrt{\pi\hbar}}} e^{-(x-y)^2\!/\hbar} \right]\,
      {1\over\pi\sqrt{2|\beta|^2-y^2}}                          \eqno(13)  $$
where appropriate factors of $\hbar$ have been restored.  Here
$\pi^{-1}(2|\beta|^2-y^2)^{-1/2}$ represents the probability for a classical
oscillator with a specified maximum amplitude $\sqrt2\,|\beta|$ to be found at
position $y$ at a randomly selected time.  Eq.$\,$(13) tells us that this
classical probability distribution is then smeared out over the width of the
quantum wave packet.  In the $\hbar\to0$ limit, the wave packet
becomes a delta function, and we recover the classical probability.

This is a rather innocuous result.  In order to discover this sort of violation
of quantum mechanics experimentally, one would have to let the oscillator run
for a long time (since the new parameter $g$ is presumably small) while keeping
precise track of the time.  On the other hand, systems such as the earth
revolving around the sun can be treated in an essentially similar way, have
been around for billions of years, and have not been seen to exhibit any
mysterious timing discrepancies.  But it is hard to translate this into a
limit on $g$ for elementary particles in any convincing way.

As another example, let us consider a free, nonrelativistic particle with unit
mass in one dimension (so that the hamiltonian is $H=\half p^2$).
(For a related discussion, see ref.\refto{marinov85}.)  Take the initial state
to be a gaussian wave packet of width $a$ and and mean velocity $p_0$:
$$\psi(x,0) = {1\over\pi^{1/4}a^{1/2}}
                   \exp\!\left[ip_0 x-{x^2\over 2a^2}\right]\;.  \eqno(14)$$
Then, at later times, we find (according to ordinary quantum mechanics)
that the probability to find the particle at position $x$ is
$$|\psi(x,t)|^2={1\over\pi^{1/2}(a^2+t^2\!/a^2)^{1/2}}\,
           \exp\!\left[-\,{(x-p_0t)^2\over a^2+t^2\!/a^2}\right]\;. \eqno(15)$$
This is easy to understand.  The initial uncertainty in position
$\Delta x\sim a$ implies an uncertainty in momentum $\Delta p\sim 1/a$.
Then, at later times, the momentum uncertainty implies an additional
uncertainty in position $(\Delta p)t\sim t/a$.  These are then compounded
quadratically to give the time dependent position uncertainty
$\Delta x(t)\sim (a^2+t^2\!/a^2)^{1/2}$ implied by eq.$\,$(15).
Meanwhile, the center of the wave packet moves with constant velocity~$p_0$,
the only possibility consistent with conservation of momentum.

Now we again consider modifying the quantum equation of motion.  Since the
hamiltonian is so simple in this case, we can take the $Q$'s to be arbitrary
functions of $p$ and still conserve energy and momentum.
Let us work out one example
in detail: a single $Q$, this time (for later mathematical convenience) equal
to the momentum $p$.  Expanding in momentum eigenstates, we see that
the probability to find the particle at position $x$ at time $t$ is
$$\langle x|\rho(t)|x\rangle={1\over2\pi}\int_{-\infty}^{+\infty}dp\,dk\,
    \widetilde\psi(p,0)
    \widetilde\psi^*(k,0)
    \,e^{i(p-k)x-i(p^2-k^2)t/2-g(p-k)^2t/2}\;.               \eqno(16)$$
where $\widetilde\psi(p,0)$ is the Fourier transform of $\psi(x,0)$:
$$\widetilde\psi(p,0)={a^{1/2}\over\pi^{1/4}}\,
                      \exp\!\left[-\half a^2(p-p_0)^2\right]\;.  \eqno(17)$$
Because the energies are continuous in this case, we cannot simply
discard the off-diagonal terms as we did for the harmonic oscillator.
Instead, we note that the choice of $Q=p$ has rendered
$\langle x|\rho(t)|x\rangle$ as a gaussian integral which can
be evaluated exactly at arbitrary times:
$$\langle x|\rho(t)|x\rangle={1\over\pi^{1/2}(a^2+t^2\!/a^2+2gt)^{1/2}}\,
     \exp\!\left[-\,{(x-p_0t)^2\over a^2+t^2\!/a^2+2gt}\right]\;.  \eqno(18)$$
We see that there is a new contribution to the uncertainty in position,
$\Delta x_{\rm new}\sim(2gt)^{1/2}$, which is compounded with
the ordinary quantum uncertainties.  The $t^{1/2}$ dependence of
$\Delta x_{\rm new}$ is characteristic of a one-dimensional random walk.
Note that, since momentum is still conserved, the center of the wave packet
still moves at constant velocity $p_0$.

This change in the nature of how wave packets spread out is again an innocuous
one.  In fact, for $g\ll 1$, the effect of $\Delta x_{\rm new}$ is negligible,
since then $gt\ll a^2+t^2\!/a^2$ at all times.

Now let us consider two free particles, with hamiltonian $H=\half(p_1^2+p_2^2)$
and initial wave function
$$\psi(x_1,x_2,0) \sim \exp\!\left[ip_{10}x_1 + ip_{20}x_2
                                   -{(x_1-x_{10})^2\over2a^2}
                                   -{(x_2-x_{20})^2\over2a^2}\right]\;.
                                                                \eqno(19)$$
We can analyze this system in the same way as before; taking $Q$ to be
equal to the total momentum $p_1+p_2$ of the two particles, we eventually find
$$\langle x_1,x_2|\rho(t)|x_1,x_2\rangle \sim
       \exp\!\left[-\,{(a^2+t^2\!/a^2+2gt)(y_1^2+y_2^2)-4gt\,y_1 y_2 \over
                    (a^2+t^2\!/a^2)(a^2+t^2\!/a^2+4gt)}\right]      \eqno(20)$$
where $y_i=x_i-(x_{i0}+p_{i0}t)$.
The $y_1 y_2$ term indicates that the presence of a second particle influences
the first.  Even though they are spatially uncorrelated initially, correlations
develop, independent of how far apart they are.  This represents a violation of
cluster decomposition.  However, for $g\ll 1$, the new correlations are small
at all times, and actually {\it decrease} with time for $t\gtwid a^2$.
Furthermore, if we trace over the states of the second particle
(integrate eq.$\,$(20) over $x_2$), we reproduce eq.$\,$(18).
Thus the second particle has no effect if we know nothing of its state.

A quantum field theory will exhibit similar behavior.  Indeed, for free field
theory this analysis can be taken over completely, except that we should use a
relativistic dispersion relation, and choose $Q$'s which respect Lorentz
covariance.  The change in the dispersion relation prevents exact evaluation of
the integrals, and even a steepest descent analysis is quite involved, but the
behavior should be qualitatively similar.  This means that loss of purity
does not force us to rethink the LSZ reduction paradigm for scattering theory,
since wave packets will remain sufficiently compact, and extra correlations
will always be small.  A more serious problem is the requirement of Lorentz
covariance, which, it turns out, is impossible to satisfy.  This is the subject
of the next section.

\subhead{3. Formal Constraints}

Since we have a theory in which correlations can appear between spacelike
separated systems, finding a Lorentz covariant version may seem to be a
hopeless task.  This is not necessarily the case: theories with tachyons,
whatever their other faults, are Lorentz covariant, and have superluminal
correlations.  Thus a reasonable question to ask is, what is the
{\it weakest possible} form of Lorentz covariance which can be demanded
of eq.$\,$(2)$\,$?

Let us begin with a concrete example of eq.$\,$(2) which {\it is} Lorentz
covariant.  Consider a scalar field $\varphi$ with lagrangian
${\cal L}=\half\partial^\mu\varphi\partial_\mu\varphi-V(\varphi)+J\varphi$,
where $J$ is a gaussian random variable with $\langle J(x)J(y)\rangle =
g\,\delta^4(x-y)$.  Here angle brackets denote an average over the probability
distribution for $J$.  This theory is manifestly Lorentz invariant, and, as
shown by BPS, leads to the evolution equation
$$\dot\r=-i[H,\r] -\half g\int d^3\!x\,\bigl[\varphi(\vec x\,),\bigl[
                            \varphi(\vec x\,),\r\bigr]\bigr]\;, \eqno(21)$$
which is a specific form of eq.$\,$(2).  This theory is not viable, since BPS
demonstrated that it produces enormous violations of energy and momentum
conservation.  It does, however, serve as an example of the form eq.$\,$(2)
could take in a Lorentz covariant framework.  We can rewrite eq.$\,$(21) as
$$\partial^0\!\r=-i[P^0,\r] -\half g\int d\Sigma^0\,
                             \bigl[\varphi(\vec x\,),\bigl[
                             \varphi(\vec x\,),\r\bigr]\bigr]\;. \eqno(22)$$
Here $\vec x$ specifies a point on the spacelike hypersurface $\Sigma$ whose
unit normal defines the time direction, and
$d\Sigma^\mu={\textstyle{1\over6}}\varepsilon^{\mu i j k}dx_i dx_j dx_k$.
Clearly, each term in eq.$\,$(22) carries a vector index in the time direction,
and no other explicit indices.  I wish to argue that this must be true of
{\it any} version of eq.$\,$(2) which can arise from an underlying, Lorentz
invariant framework.  It is by no means clear that this condition is sufficient
to ensure the existence of a full set of Lorentz transformations with the
correct algebra, but it would appear to be necessary.  This condition on
eq.$\,$(2) is what I mean by the ``weakest possible'' form of Lorentz
covariance.

We do not need to know much more about Lorentz transformations to reach a
strong conclusion.  This is fortunate:  since eq.$\,$(2) tells us that
infinitesimal time translations are {\it not} generated by the
hamiltonian~$H$, it is clearly unsafe to assume that space translations,
rotations, and boosts will still be generated by the usual
operators.\footnote{*}{In section~2, we implicitly assumed that the total
momentum operator $\scriptstyle\vec P$ generates space translations;
relaxing this may provide another way out of unpleasant conclusions
regarding nonlocal effects.}
If Lorentz invariance is to have any meaning at all, however, we must
demand that the Lorentz generators (whatever form they may take) satisfy
certain basic properties, including the following:

(1)~The total energy and momentum operators $H$ and $\vec P$
should transform as a four-vector under infinitesimal Lorentz
transformations  (so that, for example, the generator of boosts in the
$z$-direction, acting on $H$, results in $P_z$).

(2)~The infinitesimal time and space translations
$dt$ and $d\vec x$ must also transform as a four-vector.

(3)~The usual product rules should hold, so that, for example,
$M^2\equiv H^2-\vec P\,^2$ is still invariant under Lorentz transformations.

This information alone is enough to allow us to conclude that $\dot\r$
and $[H,\r]$ both transform in the same way.  I have argued that Lorentz
covariance will be lost if the final term in eq.$\,$(2) does not also have
this property.  Therefore the product of two $Q$'s must transform like $H$.
The simplest possibility is to let one $Q$ be a Lorentz scalar, perhaps
$M^2 \equiv H^2-\vec P\,^2$, and let the other $Q$ be $H$ itself.
The problem with this is that it implies a matrix $g_{\a\b}$ which is purely
off-diagonal, and which therefore has a negative eigenvalue.  One can see
the problem explicitly in free field theory.
For definiteness, take $Q_1=H$ and $Q_2=M^2$, with $g_{11}=g_{22}=0$ and
$g_{12}=g_{21}=g$.  All one-particle states
have the same value of $M^2$, and so the extra term in eq.$\,$(2)
vanishes for all one particles states, pure or mixed.
If, however, we consider a two-particle state
$\ket{p_1\,p_2}=a^\dagger(p_1)a^\dagger(p_2)\ket0$,
and evolve the initial density matrix $\r(0)=\ket{p_1\,p_2}\bra{p_3\,p_4}$,
we find at later times
$$\r(t)=
  \exp\bigl[-i(E_{12}-E_{34})t
            -g(E_{12}-E_{34})(M^2_{12}-M^2_{34})t\bigr]
               \ket{p_1\,p_2}\bra{p_3\,p_4}             \eqno(23) $$
where
$E_{ij}=E_i+E_j$,
$E_i=(\vec p\,^2_i+m^2)^{1/2}$,
$M_{ij}^2=(E_i+E_j)^2-(\vec p_i+\vec p_j)^2$,
and $m$ is the mass of one particle.  The $g$ dependent term is not
always negative; indeed its sign is frame dependent.  (The same is true
if we replace $M^2$ by some other Lorentz scalar, such as a charge.)
An off-diagonal
matrix element of $\r$ which is monotonically increasing guarantees that
$\r$ will eventually have a negative eigenvalue.  Therefore this Lorentz
covariant scheme must be rejected.

The only other possibility arises in a supersymmetric theory: then we
have the supercharges at our disposal.  They commute with the hamiltonian,
and their product has the correct transformation properties.
The appropriate version of eq.$\,$(2) is
$$\dot\r=-i[H,\r]-\half g \left[H\r+\r H
-2\bigl(\sbar^{\,0}\bigr)^{\dot\a\a}\bigl(\Qad\r\Qa+\Qa\r\Qad\bigr)\right]\;.
                                                      \eqno(24)$$
$\Tr\dot\r=0$ follows from the supersymmetry algebra
$\bigl\{\Qa,\Qad\bigr\}=\bigl(\s^\mu\bigr)_{\a\dot\a}P_\mu$.
(We may use conventions in which both
$\bigl(\s^0\bigr)_{\a\dot\a}$ and $\bigl(\sbar^{\,0}\bigr)^{\dot\a\a}$
are equal to $\d_{\a\dot\a}$.)  Eq.$\,$(24) also conserves energy and momentum,
since $\Qa$ and $\Qad$ commute with $H$ and $\vec P$.

The problem with eq.$\,$(24) is that it does not conserve angular momentum.
Although it is straightforward to show (using the transformation properties
of $\Qa$ and $\Qad$) that while $\Tr\vec J\r$ remains constant,
$\Tr\vec J\,^2\r$ does not.  It is easy to see this physically:
$Q\sim b^\dagger a$, where $b^\dagger$ creates a fermion and $a$ destroys
a boson.  An initial state of one boson at rest will be converted to a state
of one fermion at rest by eq.$\,$(24).  This fermion is equally likely to have
spin up or spin down, and so $\Tr\vec J\r$ remains zero, but clearly
$\Tr\vec J\,^2\r$ has changed.  Thus this scheme must be rejected
as well, and we are forced to conclude that eq.$\,$(2) is incompatible with
the weakest possible form of Lorentz covariance.

This objection would not arise in two dimensions, where there is
no angular momentum.  Perhaps eq.$\,$(24) has some
role to play on the superstring worldsheet.

\subhead{4. A possible loophole}

The conclusion just reached---that the evolution of pure to mixed states is
incompatible with the weakest possible form of Lorentz covariance---was
based on the assumption that the $Q$'s in eq.$\,$(2) must commute with $H$.
BPS showed that this is the only possibility if $g_{\a\b}$ is a real symmetric
matrix with nonnegative eigenvalues.
We must still consider the possibility that $g_{\a\b}$ does not obey
these conditions (since positivity alone is a sufficient
condition to avoid development of negative probabilities,
and even that may not be necessary).  Is it possible to conserve
energy, momentum, and angular momentum, even if the corresponding
operators do not commute with the $Q$'s? Surprisingly, the answer is yes.
The example I have found is not a viable theory because it allows $\r$ to
develop negative eigenvalues; however, I have been unable to prove that a
model without this fatal flaw does not exist.  This leaves open a small
chance to maintain the weak form of Lorentz covariance along with the
familiar conservation laws.

The simplest theory which conserves energy even though $[H,Q]\ne0$
is based on the hamiltonian $H=\bd b$, where $b$ is a fermion operator
obeying the usual anticommutation relations $\{b,b\}=0$ and  $\{\bd,b\}=1$.
We take $Q_1=b^\dagger+b$ and $Q_2=i(b^\dagger-b)$, with
$g_{11}=-g_{22}=g$ and $g_{12}=g_{21}=0$.
Obviously, $g_{\a\b}$ has a dangerous negative eigenvalue.  Eq.$\,$(2) becomes
$$\dot\r=-i[H,\r] - 2g\bigl(\bd\r\bd+b\r b\bigr)\;.            \eqno(25)$$
We can now see why energy is conserved; if we compute $\Tr H^n\dot\r$,
all the extra terms vanish because we always get a factor of either $b^2=0$
or $b^{\dagger 2}=0$.  However, it is not hard to show that, in general, an
initially pure $\r$ immediately evolves into one with a negative
eigenvalue.\footnote{*}{After this paper was circulated as a preprint,
Jun Liu found a modified version of eq.$\,$(25) which preserves the
positivity of $\r$\refto{liu93}.}

The model may be made less trivial by extending
it to a set of independent operators $\bi$, with $H=\sum_i\bid\bi$:
$$\dot\r = -i[H,\r]-2g\,{\ts\sum\limits_i}\,
                  \bigl(\bid\r\bid+\bi\r\bi\bigr)\;. \eqno(26)$$
It is easy to check that we still have $\Tr H^n\dot\r=0$.  This version
can be promoted to a field theory of free (Majorana) fermions:
$$\dot\r = -i[H,\r]
       -2g\,{\ts\sum\limits_s}\int d^3p\,\bigl[b^\dagger(p,s)\r b^\dagger(p,s)
                                    +\hbox{h.c.}\bigr]\;,       \eqno(27)$$
where $b^\dagger(p,s)$ creates a fermion with momentum $p$ and
spin $s$.  The extra term has the correct Lorentz transformation
properties, but is nothing very simple (or local) in position space.
And, as before, it leads to negative probabilities.

This existence of this example is quite annoying, since it prevents a general
proof of the necessity of $[H,Q]=0$ for energy conservation, yet is both
untenable (because of the negative probabilities) and extremely contrived.
It is hard to imagine a physical process (such as wormholes) leading to
anything as complicated in position space as eq.$\,$(27).

\subhead{5. Conclusions}

A reexamination of the question of whether or not pure states can evolve into
mixed states shows that phenomenological problems associated with loss of
locality are less severe than previously believed.  However, it seems
impossible to maintain both Lorentz covariance of the modified evolution
equation and the familiar conservation laws.  A loophole in the general
argument leading to this conclusion was noted, but the only example I could
find which sneaks through it is rather special, and in any case results in
negative probabilities.  The fascinating possibility that purity may not be
eternal is still out of reach.

I would like to thank two anonymous referees for lively correspondence and
helpful suggestions which substantially improved this paper.  This work was
supported in part by NSF Grant No.~PHY--86--14185.

\vfill\eject
\references
\baselineskip=16pt

\refis{hawking83}S. W. Hawking, \cmp, 87, 1983, 395. 

\refis{marinov74}M. S. Marinov, \journal JETP Lett., 15, 1972, 479;
\journal Sov. J. Nucl. Phys., 19, 1974, 350.

\refis{gross84}D. J. Gross, \np, B236, 1984, 349.   

\refis{ehns84}J. Ellis, J. S. Hagelin, D. V. Nanopoulos, and M. Srednicki,
\np, B241, 1984, 381.  

\refis{hawking84}S. W. Hawking, \np, B244, 1984, 135. 

\refis{bps84}T. Banks, M. E. Peskin, and L. Susskind, \np, B244, 1984, 125.

\refis{gs88a}S. B. Giddings and A. Strominger, \np, B306, 1988, 890.

\refis{gs88b}S. B. Giddings and A. Strominger, \np, B307, 1988, 854.

\refis{coleman88a}S. Coleman, \np, B307, 1988, 867.

\refis{ksb89}I. Klebanov, L. Susskind, and T. Banks, \np, B317, 1989, 665.

\refis{marinov85}M. S. Marinov, \np, B253, 1985, 609.

\refis{liu93}Jun Liu, Stanford Univ. preprint SU-ITP-93-1, hep-th/9301082.

\endreferences\endit\end